\documentclass[draft]{agujournal2019}
\usepackage{url} 
\usepackage[inline]{trackchanges} 
\usepackage{soul}
\usepackage{amsmath}
\usepackage{amsfonts}
\usepackage{amssymb}
\usepackage{amsthm}
\usepackage{physics}
\usepackage[english]{babel}
\usepackage{fontenc}
\usepackage{graphicx}
\usepackage{natbib}

\let\oldequation\equation
\let\oldendequation\endequation
\renewenvironment{equation}
  {\linenomathNonumbers\oldequation}
  {\oldendequation\endlinenomath}
\let\oldalign\align
\let\oldendalign\endalign
\renewenvironment{align}
  {\linenomathNonumbers\oldalign}
  {\oldendalign\endlinenomath}

\draftfalse

\begin{document}

%
%

\title{Correlation based modeling \\
of the ionospheric magnetic field}

%
%




\authors{K. Ferrat\affil{*, 1}, M. Schanner\affil{*, 1, 2}, V. Lesur\affil{3}, M. Holschneider\affil{1}, \\
C. Stolle\affil{4}, J. Baerenzung\affil{1}}

\affiliation{*}{Equal contribution}
\affiliation{1}{University of Potsdam}
\affiliation{2}{GFZ Potsdam}
\affiliation{3}{IPG, Paris}
\affiliation{4}{IAP, Kühlungsborn}

\newcommand\matthias[1]{{\color{red}[matthias: #1]}}
\newcommand\vincent[1]{{\color{red}[vincent: #1]}}
\newcommand\khadidja[1]{{\color{red}[khadidja: #1]}}
\newcommand\claudia[1]{{\color{red}[claudia: #1]}}
\newcommand\boris[1]{{\color{red}[boris: #1]}}
\newcommand\julien[1]{{\color{red}[julien: #1]}}




\correspondingauthor{Khadidja Ferrat}{ferkadi@icloud.com}




\begin{keypoints}
\item Correlation  based modeling
\item Empirical covariance structure
\item Monitoring of Ionosphere
\end{keypoints}

%
%

%
%

\begin{abstract}
\noindent	Patterns of the magnetic signature of ionospheric currents, generated from an empirical model based on satellite observations, are used to build a statistical correlation based model for ionospheric fields. In order to stabilize the dynamics and to take into account the dominant role of the sun, the fields are represented in solar magnetic coordinates.
The covariance structure is analyzed and a second order process that approximates the full dynamics is generated.
We show that for synthetic data observations located at the Earth observatories, the full ionospheric field pattern as observed on the earth surface can be reconstructed with good precision. As a proof of principle we provide a first application to the inversion based on real observatory data.
\end{abstract}


%
%

%


%
%
%
%

\section{Introduction} \label{sec:1}
Geomagnetic field records near the Earth's surface are the combination of contributions from numerous sources. The dominant source is the core magnetic field, but other significant contributions come from the lithosphere, the ionosphere, the magnetosphere and field-aligned currents linking the two latter regions 
\citep[e.g.][]{Alken:2020}. 
Weaker contributions are due to field induced in the conductive parts of the Earth (mainly in the mantle, the crust and the Oceans), due to Ocean currents and tides, and probably due to other contributions that are not well identified. Between all these contributions, the magnetic field generated in the ionosphere is one of the most difficult to describe accurately. 
This is firstly, because on the day-side of the Earth the systems of currents generating the magnetic field flow in the ionosphere at relatively low altitudes -- typically $110$km. As a result, the recorded signals on the Earth surface are at relatively small spatial wavelengths \citep{Prokhorov:2018}, and the sparse network of magnetic observatories (or variometer stations) gives only limited information on the ionosphere magnetic field strength, orientation and variations. Furthermore, the structure of these systems of currents is constrained, on one hand, by the shape of the core magnetic field that rotates together with the Earth and, on the other hand, by the position of the Sun relative to the Earth that varies not only over a single day, but also over the year. Because of this combination, even the average ionospheric magnetic field cannot be parametrized in a simple manner. Finally, the ionosphere responds to the fluctuation of its environment, such as the energy input through the field aligned current, the gravity waves, tides, the particle flux from the Sun, and numerous other perturbations that generate variations of the ionospheric magnetic field particularly difficult to describe \citep{Yamazaki:2017a}.\\
The general structure of the aforementioned system of currents, that generate a magnetic field which can be recorded at the Earth's surface, has been known for some time. All currents are situated on the day-side of the Earth and present a significant local-time variability. At mid latitudes, there is the large scale Solar quiet (Sq) vortex \citep{Yamazaki:2017a}, flowing anti-clockwise (resp. clockwise) in the Northern (resp. Southern) hemisphere, and source of a signal reaching some $20 \mbox{nT}$ at the Earth's surface. Along the magnetic equator the electrojet and counter electrojet systems flow East- and West-ward respectively. They are generating relatively small scale magnetic fields, and studies based on satellite data have shown their dependence upon atmospheric tides \citep{Yamazaki:2017b}. At high latitudes, the patterns of currents are highly variable, responding rapidly to changes in the magnetospheric activity. They are typically described through two types of currents: Hall and Pedersen currents \citep{Newell:1968}, the former contributing mainly to the polar electrojet. This system of currents is the most significant contribution at high latitudes to magnetic observations on Earth's surface. Its shape and strength depend significantly on the orientation of the interplanetary magnetic field (IMF) and the solar wind speed. The polar electrojet extends towards mid-latitudes during {geo\-mag\-netic} storms. There are numerous other smaller scale, possibly higher altitude, currents generating less distinctive signatures at Earth's surface and the simplified view of the ionosphere system of currents is not precise enough to allow a proper description of the rapidly varying and complex observed signals at the Earth's surface.\\
To help the description of the ionospheric magnetic field, it is possible to rely on empirical or semi-empirical models of the ionosphere based on different types of information (winds, electron contents, ionosphere glow, etc). There are also theoretical models, solving the magneto-hydrodynamic equations for the ionosphere. Three of these are the General Circulation Model (GCM,  \citealt{Dickinson:1981,Qian:2014}), the Upper Atmosphere Model (UAM, \citealt{Namgaladze:2017}) and components of the Space Weather Modeling Framework (SWMF, \citealt{SWMF}). However, traditionally the geomagnetic models describing the signals generated by the ionosphere do not use information from these models. As an example, the Comprehensive Model (CM) of the Earth magnetic field \citep{Sabaka:2020} has a part dedicated to the ionospheric contribution. The ionosphere signal is described through a single map of currents, at $110$ km altitude. The system's geometry is constrained by the Apex coordinate system \citep{Laundal:2017}. The map of currents, and therefore the magnetic field, is described up to spherical harmonic 60, and the limited information provided by observatory data (and more recently by satellite data) is completed by imposing symmetries and smoothness to the model, as well as strict temporal periodicities. The same approach has been used in the ionosphere dedicated L2 products of the Swarm mission \citep{Chulliat:2016, Jin:2019}.\\
In this work, our aim is not to give information on the physical processes inside the ionosphere, but as a proof of concept, to reconstruct anywhere close to the Earth's surface the magnetic field generated in the ionosphere. The data used to achieve this goal are solely mean observatory vector magnetic data. To complement the globally sparse data, we use a correlation based technique \citep{Holschneider:2016}, where the prior information on the spatial characteristic of the magnetic field generated in the ionosphere relies on statistics extracted from empirical models based on satellite data. As described below, we do not apply pre-processing to the observatory data to correct for non-ionospheric contributions. We rather co-estimate in a strongly simplified way all the magnetic field components together with the one due to the ionosphere.\\
The next section is the main part of the paper and describes the theoretical and mathematical framework that underpin our modeling approach. The viability of the technique is then tested in Section~\ref{sec:closedloopsim} through a closed loop process using a synthetic data set. Finally, the whole process is applied to real observatory data. This is described in Section~\ref{sec:obsdatainverson}. Section~\ref{sec:conclusion} concludes the paper.

\section{Modeling framework}
In \cite{Holschneider:2016} the authors proposed to use prior covariance kernels for the geomagnetic field components to build models of magnetic fields from observations and separate the various components. By assimilating night time data only, they could neglect the influence of the ionospheric field and therefore avoid modeling it directly, treating it as a nuisance component. As a proof of concept, we propose the opposite in this paper: We construct a correlation based prior for the ionosphere and treat the other fields in a very simplified fashion. A more extensive study could then join the methods and extend the approach of \cite{Holschneider:2016}. As an introduction, we briefly recap the correlation based inversion in the following:\\
We assume that our field model $\Psi(x)$ is a~priory characterized by a Gaussian process with mean $\mu_\Psi$ and covariance $\Gamma_\Psi$:
\begin{equation}
	\Psi\sim\mathcal{N}(\mu_\Psi,\Gamma_\Psi)
\end{equation}
Thus
$$
\mathbb{E}(\psi(x)) = \mu_\Psi(x),\quad \mathbb{V}(\Psi(x), \Psi(x^\prime)) = \Gamma_\psi(x, x^\prime)
$$
The measurements are given by noisy observations of some linear functional $L_k$, $k=1,\dots,K$ of  
$\Psi$. The observation equation is therefore
$$
L_k\psi = l_l + \epsilon_k, \quad \epsilon_k \sim N(0, \Sigma)
$$
The measurement noise $\epsilon$ is supposed to be Gaussian. It may be correlated Gaussian noise of zero mean and covariance $\Sigma$.
These observations $l_k$ are then multivariate Gaussian variables
\begin{align*}
	\mathbb{E}(l_k) &= \mu_{l_k} = L_k\mu_\Psi \\
	\mathbb{V}(l_k,l_{k^\prime}) &= \Gamma_{k, k^\prime}  + \Sigma_{k, k^\prime}=
L_k\Gamma_\Psi L_{k^\prime}^T + \Sigma_{k, k^\prime}\\
	\mathbb{V}(\Psi(x), l_k) &= \Gamma_{\Psi(x),l} = (\Gamma_\Psi L^T_k)(x).
\end{align*}
The posterior distribution can be derived by standard Bayesian calculus. The posterior is also a Gaussian process and its mean and covariance are given by ($\boldsymbol{l}=[l_1,l_2\dots]^T$)
\begin{align}
    \mathbb{E}(\Psi |\boldsymbol{l}) &= \mu_\Psi + \Gamma_{\Psi, \boldsymbol{l}} (\Gamma_{\boldsymbol{l}, \boldsymbol{l}} + \Sigma_{\boldsymbol{l}, \boldsymbol{l}})^{-1} (\boldsymbol{l} - \mu_{\boldsymbol{l}} )\\
\mathbb{V}(\Psi | \boldsymbol{l}) &= \Gamma_\Psi  -  \Gamma_{\Psi, \boldsymbol{l}} (\Gamma_{\boldsymbol{l}, \boldsymbol{l}}+\Sigma_{\boldsymbol{l}, \boldsymbol{l}})^{-1} \Gamma_{\Psi, \boldsymbol{l}}^T
\end{align}

Observe that in the limit of zero observational noise, the actual observations $\boldsymbol{l}$ enter only in the equation of the posterior expectation through the deviation between the predicted mean and its observed counterpart. The posterior covariance is independent from the actual value of the observations. Only the measurement geometry of the observables $L$, expressed through their prior covariance structure, enters here.  However, for noisy observations the values of the observations do enter into the posterior covariance. 

These formulas can be easily carried over to the situation, where the fields become function of space and time $\psi=\psi(x,t)$ as well as their prior covariance structure. 

\subsection{Correlation patterns of Ionospheric data}\label{sec:corrpattern}

\begin{figure}
	\begin{centering}
		\includegraphics[width=\linewidth]{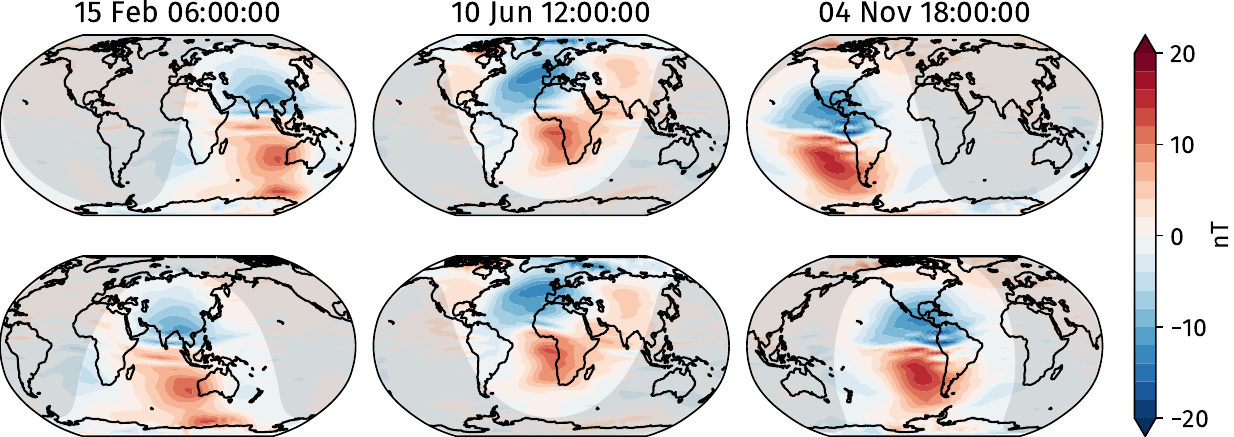}
	\end{centering}
	\caption{Geocentric (top) and solar magnetic (bottom) downward component of the ionospheric field at the Earth's surface for different epochs throughout the year. The night-side is indicated by the shaded area.}\label{Fig:ionosphere-snapshot}
\end{figure}

In the source free region between the Earth's surface and the lower boundary of the ionosphere, the ionospheric field can be completely determined by a time dependent potential $\Phi_\text{iono}(x,t)$ through the relation
\begin{equation}
    B_\text{iono} = -\nabla \Phi_\text{iono}.
\end{equation}

A large part of the dynamics of the ionospheric fields as observed on the surface of the Earth is due to the relative position of the sun to the point of observation.  In order to get rid of this trivial component, we seek to produce a statistical model of the ionospheric field in Solar Magnetic (SM) coordinates \citep{Laundal:2017}.  In this coordinate system, the movement of the sun is largely reduced and the ionospheric field patterns stabilize. If  $x =[x_1, x_2, x_3] \simeq [r, \theta,\phi]$ denotes  the Euclidean, respectively spherical coordinates of a point at time $t$ in a geocentric coordinate system then its coordinates in the SM system,  $x^\text{SM} = [x_1^{SM}, x_2^{SM}, x_3^{SM}] \simeq [r^\text{SM}, \theta^\text{SM}, \phi^\text{SM}]$ are obtained through a mapping $\Omega_t$
\begin{equation}
	\Omega_t(x) = x^\text{SM} .
\end{equation}
In Euclidean coordinates, $\Omega_t$ is a rotation, in the spherical coordinates however it is a non-linear map. Then, we propose to describe the ionospheric field in SM coordinates as a multivariate space time Gauss process which may be projected back to the geocentric system with the help of $\Omega_t$. 
In Cartesian coordinates (for the position and the field) we simply have
$$
B(x,t) = \Omega^T_t B^{SM}(\Omega_t(x), t)
$$

Suppose in the SM system the ionosphere is described through an average field $\mu^\text{SM}(x)$ and correlated fluctuations of the ionospheric field around $\mu^\text{SM}$.
In terms of the potential, the mean is given by
\begin{equation}
	\mathbb{E} \qty(\Phi^\text{SM}(x^\text{SM}, t)) = \mu^\text{SM}(x^\text{SM}, t)
\end{equation}
and for two points $x^\text{SM}$ and $y^\text{SM}$ the covariance reads
$$
\mathbb{V} (\Phi^\text{SM}(x^\text{SM}, t),\Phi^\text{SM}(y^\text{SM}, \tau)) =
\Gamma^\text{SM} (x^\text{SM}, t, y^\text{SM}, \tau).
$$
In the geocentric system we obtain for two points $x$, $y$ and a time
$t$ and $\tau$:
\begin{align*}
	\mathbb{E} ( \Phi_t(x,t) ) &= \mu(x,t) = \mu^\text{SM}( \Omega_t(x) ), \\
	\mathbb{V} (\Phi(x, t)\Phi(y, \tau)) &= \Gamma(x,t, y, \tau) =
\Gamma^\text{SM}(\Omega_t(x), t ,\Omega_\tau(y), \tau)
\end{align*}
It is convenient to expand the potential $\Phi$ in terms of the spherical harmonic functions $Y_{\ell, m}$ with time dependent Gauss coefficients. We write in the SM coordinate system:
\begin{equation}
	\Phi^{SM}(x^\text{SM}, t) =	\sum_{\ell, m} g_{SM, \ell}^m(t)
	Y_{\ell, m}(\hat x^\text{SM}) \left(\frac{|x^\text{SM}|}{R}\right)^\ell
\end{equation}
$\hat x^\text{SM}$ is the projection of the position $x^\text{SM}$ to the unit sphere and $|x^\text{SM}| = |x|$ is the length of the position vector. The $g_{SM,\ell}^m$ are the Gauss coefficients in the SM coordinate system. 
In the Earth fixed coordinate system the potential can be written as
$$
\Phi(x, t) = \Phi^{SM}(\Omega_t(x), t) = \sum_{\ell, m} g_{SM, \ell}^m(t)
	Y_{\ell, m}(\Omega_t \hat x) \left(\frac{|x|}{R}\right)^\ell
$$
It is possible to use explicit kernel functions to model the a priory 
covariance structure of $\Phi$. We shall use an empirical covariance structure obtained 
from simulations. It shall be described through a multi-variate distribution of the Gauss coefficients in the SM system.

\subsection{Empirical covariance structure}\label{sec:empiricalcovariance}

\begin{figure}
	\begin{center}
		\includegraphics[width=\linewidth]{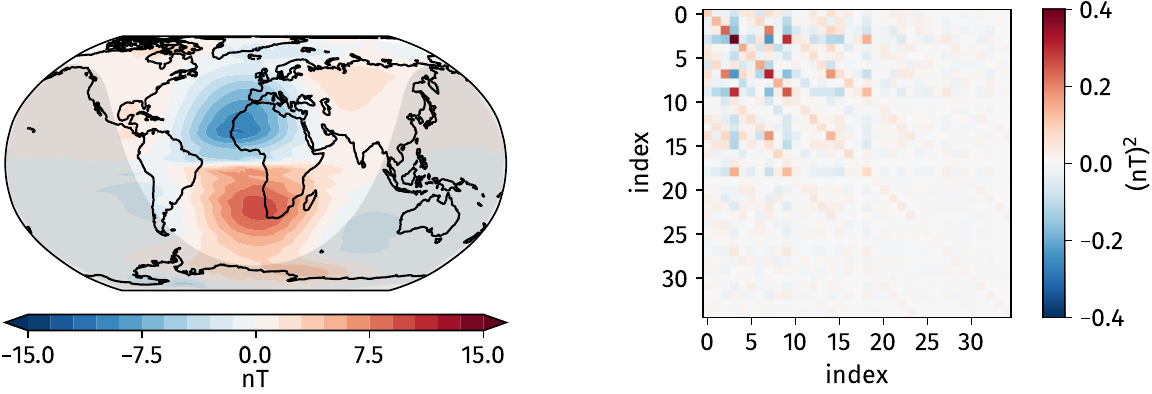}
		\caption{Visual representation of the statistical properties derived from the one year simulation of the ionospheric model.
		Averaged downward component of the ionospheric field evaluated at the Earth's surface and projected in the SM coordinate system (left). Covariance matrix associated with the Gauss coefficients of the field (right). ``index'' refers to the Gauss coefficient index in the standard order ($g_1^0$, $g_1^1$, $g_1^{-1}$, $g_2^0$, $g_2^1$, $g_2^{-1}$, $g_2^2$,...)}
		\label{fig:mean_and_cov}
	\end{center}
\end{figure}

As a proof of concept, correlation based inversion is applied to the ionospheric magnetic field. The first step in this endeavour consists of constructing a prior model. As explained in the previous section, this consists of constructing a covariance matrix for the ionosphere Gauss coefficients $g_\ell^m$ in the SM coordinate system. The aim is to reconstruct the field from observatory data only. The prior covariance structure will be inferred from a global model of the ionosphere. Originally, the idea was to use a theoretical model based on magnetohydrodynamics (MHD). The initial choice was the UAM model \citep{Namgaladze:2017}, which was extended by a Biot-Savart integrator to calculate magnetic fields of ionospheric origin at the Earth's surface \citep{Prokhorov:2018}. However, due to geopolitical reasons the relevant model output is no longer available. To our knowledge, no other MHD model (such as SWMF) contains a Biot-Savart integrator for the ionosphere. Therefore, we resort to an empirical model based on observatory and satellite data. The model we choose is the DIFI-7 model, or SWARM L2 Product \texttt{MIO\_SHA}. This model is constructed by applying a modification of the comprehensive modeling chain \citep{Chulliat:2013, Sabaka:2020} and reported as a set of Gauss coefficients up to degree 60 and order 12, decomposed into periodic contributions. From this set of coefficients, the model can output the ionospheric contributions to the magnetic field at the Earth's surface for an arbitrary point in space and time and for varying solar conditions, parameterized by a linear dependence on the solar flux index F10.7. According to its specifications, the model is valid between quasi-dipole latitudes -55$^\circ$ and 55$^\circ$. To be able to replace the empirical component by an indepently derived one in the future, we do not use the DIFI-7 Gauss coefficients directly, but take a `detour' over the magnetic field output in the following way:

For an entire year, hourly snapshots of the ionospheric magnetic field are generated with the DIFI-7 model, assuming constant solar conditions to be independent of the F10.7 index. An example of three snapshots is displayed in Figure~\ref{Fig:ionosphere-snapshot}, where the downward component of the field $B_{\text{iono}; Z}$ is represented on the Earth's surface in the geocentric (top) and the solar magnetic (bottom) coordinate systems. For each snapshot $t$, $B_{\text{iono};Z}(t)$ is evaluated on a Gauss-Legendre grid in SM coordinates of radius $R=6731.2$km (the mean radius of the Earth). The field can then be transformed into spherical harmonics coefficients $g_{\ell}^m$ through the relation
\begin{equation}
g_{\ell}^m(t) = \sum_k w_k\, \alpha_\ell \, B_{\text{iono};Z}(x^\text{SM}_k, t))
Y_{\ell, m}( \hat x^\text{SM}_k)~,
\end{equation}
where $w_k$ are the weights associated with each grid point $x^\text{SM}_k$, $Y_{\ell,m}$ are the (real valued) Schmidt semi-normalized spherical harmonics and $\alpha_\ell$ is a normalizing factor. The maximum degree of the spherical harmonics expansion was set to $\ell_{max}=40$ and the maximum order was kept at $40$ as well, to differ from the DIFI-7 model. In future studies, this parametrization may change. The set of hourly snapshots for a whole year consists of $1680$ coefficients for each of the $8760$ hours in the year. From this dataset, the random model of the ionosphere based on the empirical covariance structure can now be estimated in terms of the correlation structure of the $g_\ell^m(t)$:
\begin{align*}
	\mu_{\ell, m} &= \frac{1}{N}\sum_{n=0}^{N-1} g_\ell^m(t_n)\\
	\Gamma_{\ell,m; \ell^\prime,m^\prime} &= \frac{1}{N-1} \sum_{n=0}^{N-1} (
g_\ell^m(t_n) - \mu_{\ell,m} ) (g_{\ell'}^{m'}(t_n) -
\mu_{\ell^\prime,m^\prime} )
\end{align*}
$t_n$ are the hours in the year. From these quantities the prior mean field at any point can be computed (in Cartesian coordinates) through
\begin{align*}
	\mu_B(x; t) &= -\nabla  \sum_{\ell,m}  \mu_{\ell,m}\, Y_{\ell,m}( \Omega_t \hat x )
	\frac{|x|^{\ell}}{R^{\ell}}\\
	&= - \Omega_t \sum_{\ell,m}  \mu_{\ell,m}\, (\nabla \tilde Y_{\ell,m})( \Omega_t \hat x )
	,\quad \tilde Y_{\ell,m}(x)  = Y_{\ell, m}(\hat x) \frac{|x|^{\ell}}{R^{\ell}}
\end{align*}
The prior covariance of the ionospheric magnetic field at $x$ and $y$ in the source free region below the ionosphere is given by
\begin{equation}
	\label{eq:priorcovarformula}
	\begin{aligned}
		\Gamma_B(x,y; t)  &=  -\nabla \sum_{\ell,m} \sum_{\ell^\prime,m^\prime}
		Y_{\ell,m}( \Omega_t \hat x) \Gamma_{\ell,m;\ell^\prime,m^\prime} Y_{\ell^\prime,m^\prime}( \Omega_t \hat y)
		\frac{|x|^{\ell}|y|^{\ell^\prime}}{R^{\ell+\ell^\prime}} \nabla^T\\
		&= - \sum_{\ell,m} \sum_{\ell^\prime,m^\prime}
		\Omega_t\nabla \tilde Y_{\ell,m}( \Omega_t \hat x) \Gamma_{\ell,m;\ell^\prime,m^\prime}
		\tilde Y_{\ell^\prime,m^\prime}( \Omega_t \hat y) \nabla^T \Omega^T_t
	\end{aligned}
\end{equation}
Note that for each pair of points $x$, $y$, the covariance $\Gamma_B(x,y; t)$ is a $3\times3$ matrix since it contains the correlation of all field components at $x$ and at $y$. In figure~\ref{fig:mean_and_cov} we depict the average ionospheric field on the Earth's surface as extracted from the DIFI-7 model.

\begin{figure}
	\begin{centering}
		\includegraphics[width=\linewidth]{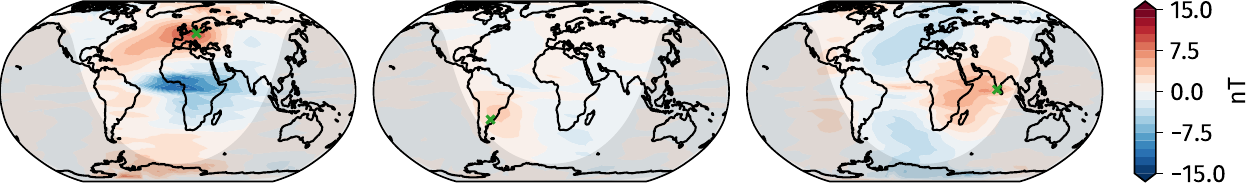}
	\end{centering}
	\caption{Correlation structure derived from the DIFI-7 ensemble for three locations, marked by a green X. From left to right: Potsdam (Germany), Buenos Aires (Argentina) and Kochi (India) at 12 UTC. The correlation pattern is not rotation invariant, but depends on the relative position of the sun.}\label{fig:covar-maps}
\end{figure}

The full spatial correlation structure is difficult to visualize. However, for a fixed point $x$ and time $t$ we can plot $y\mapsto \Gamma_t(x,y)$ as a function of $y$. This is done for $3$ different locations in figure~\ref{fig:covar-maps}. As you can see, the prior correlation structure is not rotation invariant. It incorporates the dynamical correlation patterns obtained from the DIFI-7 model simulations.

\section{Closed loop simulation of inversion}
\label{sec:closedloopsim}
In a first step, we perform a closed loop simulation to evaluate the capacity of our method to recover the ionospheric surface field from observatory data alone. To do so, we generate artificial observations at the locations of the observatories $x_i$, $i=1, \dots, N_\text{obs}$ using the DIFI-7 model. The DIFI-7 model is valid between quasi-dipole latitudes -55$^\circ$ and 55$^\circ$. Data from observatories located outside of this region are excluded in the inversion. To these artificial observations we add artificial measurement noise, randomly drawn from the Gaussian distribution $\mathcal{N}(0,\sigma^2\mathbb{Id}$). $\sigma$ was set to 3~nT. The inversion is based on a Bayesian technique, where the prior distribution of the field is characterized by the empirical covariance structure estimated as described in the previous section. The full covariance matrix of the observations is then the sum of the model prior covariance and the measurement noise
$$
 \Sigma(x_i, x_j) = \Gamma_{i,j} + \sigma^2 \mathbb{Id}
$$
For simplicity, we assume Gaussian white noise on all components of the observed magnetic field. From these observations, the posterior mean at any point $y$ on the surface of the earth can be computed as follows
$$
\mathbb{E}(B(y; t) | B_{obs} ) = \mu(y) + \sum_{i}\Sigma(y, x_i)  ( \Sigma^{-1} ( B_{obs} - \mu ) )_{i}
$$
The posterior covariance then reads
$$
\mathbb{V} ( B(x; t), B(y; t)  | B_{obs} ) =  \Gamma_{x,y} - \sum_{i,j}  \Gamma_{x,i} \Sigma^{-1}(x_i, x_j) \Gamma_{j,y}
$$
\begin{figure}
	\begin{center}
		\includegraphics[width=\linewidth]{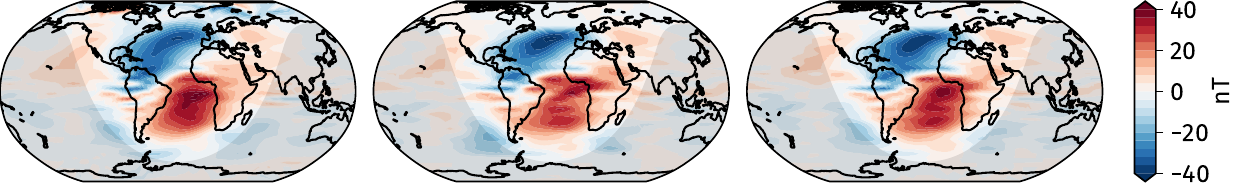}
		\label{fig:closedloop}
		\caption{Downward component of the synthetic ionospheric field (left) and its reconstruction (middle and right), expressed at the Earth's surface in SM coordinate. The right reconstruction uses free modes for the mean and the first four EOFs. See the text for details.}
	\end{center}
\end{figure}

\begin{figure}
	\begin{center}
		\includegraphics[width=\linewidth]{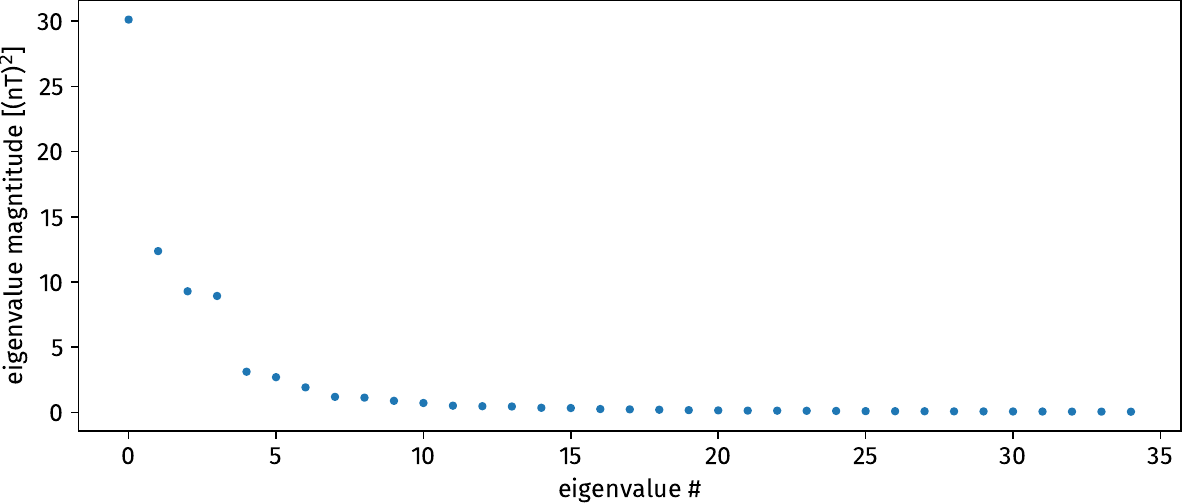}
		\caption{Eigenvalues of the covariance matrix derived from the DIFI-7 ensemble. The first dominant eigenvalue clearly sticks out from the others, which decay rapidly.}
		\label{fig:spectrum}
	\end{center}
\end{figure}

\begin{figure}
	\begin{center}
		\includegraphics[width=0.8\linewidth]{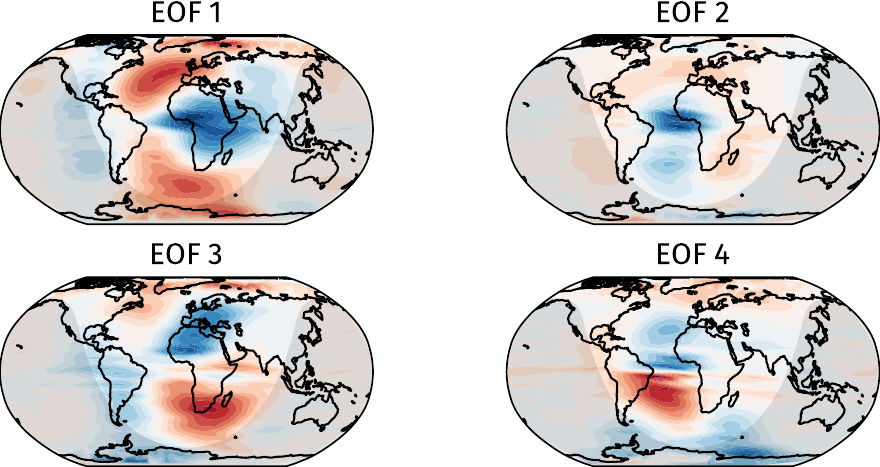}
		\caption{Downward component of the first $4$ EOFs, derived from the covariance structure of the DIFI-7 ensemble.}
		\label{fig:eof}
	\end{center}
\end{figure}

As seen in figure~\ref{fig:closedloop}, the mid-latitude solar quiet ionospheric field can be reconstructed from artificial measurements at observatory locations quite well. This shows that in principle, if the ionospheric field could be observed separately at the observatory locations, it could be reconstructed, at least to the degree of dynamical details that are contained in the DIFI-7 model. The main reason for this seems to be the effective dimension of the ionospheric field at the surface of the Earth. A possible way of quantifying the effective dimension is to analyze the spectral decay of the eigenvalues associated with the covariance matrix. For this we have performed an eigenvalue decomposition of $\Gamma$. Figure~\ref{fig:spectrum} depicts the spectrum, and as evident the eigenvalues decays very rapidly. Therefore the whole variability of the ionosphere is essentially characterized  by the first few eigenfunctions (EOF). In Figure~\ref{fig:eof}, the downward component associated with the first $4$ EOFs are displayed. We propose to add additional free modes, associated with the four largest eigenvalues, to the inversion model. More details are given in the following section. An improved reconstruction of the synthetic data is depicted in the right panel figure~\ref{fig:closedloop}.

\section{Application to observatory data}
\label{sec:obsdatainverson}
We now show how to use observatory data to infer the state of the ionosphere. The empirical DIFI-7 model of the ionosphere has served to construct a prior covariance model of the ionosphere as well as the covariance between the magnetic field at the location of the observatories and the ionospheric component. This allows us to separate this external ionospheric component from all other components through its specific covariance properties. One of the difficulties of observatory data resides in the fact that the data coverage is very heterogeneous in space and not necessarily continuous in time. Luckily, within the correlation based modeling such heterogeneous data can be dealt with in a straight forward way. We will use a highly simplified approach where we model the observed field at the observatories as sum of three a\,priory independent components
\begin{equation}
B_\text{obs} = B_\text{iono} + B_\text{const} + B_\text{noise}.
\end{equation}
The field $B_\text{const}$ is a component that on the considered time scale of one day is assumed to be constant. It contains internal and external components. In particular, we do not take into account an explicit model for the secular variation. The constant field for each observatory  contains the crustal field, the main field, a baseline offset, which is individual for each observatory, and an average external component (containing slowly varying magnetospheric field contributions)
\begin{equation}
B_\text{const} = B_\text{main} + B_\text{crust} + B_\text{offset} + B_\text{mean-ext}.
\end{equation}
We do not attempt to separate these components here. Moreover, we do not consider the a\,priori correlations present in the main field components of nearby observatories. The noise term $B_\text{noise}$ is assumed to be uncorrelated Gaussian noise, with individual variances for each epoch and location.
\begin{equation}
	B_\text{noise} \sim \mathcal{N}(0, \Sigma)~,
\end{equation}
where $\Sigma$ is a diagonal matrix. The ionospheric field covariance is constructed from a spatial part, introduced in the previous section, together with a Mat\'ern covariance for the temporal part. The correlation time is fixed to ten minutes, to regularize the inversion.
As mentioned in the previous section, we allow some additional degrees of freedom for the ionospheric field. The a\,priory mean and the first four EOFs are considered as free modes with a flat prior. The motivation for this is that during the derivation of the a\,priory covariance model, the solar activity was not considered. By including the free modes, we allow the model to scale the essential dynamics according to the data. This corresponds to the assumption that although the activity level changes, the correlation structure (i.e. normalized covariance) remains the same. To summarize we model the ionosphere in the SM system as
\begin{equation}
	B_\text{iono}^{SM}(x,t) = \beta_{0,t} \mu_B^{SM}(x) + \sum_{i=1}^4\beta_{i,t} E\!O\!F_i(x) + \eta(x,t)
\end{equation}
and $\eta$ has the covariance
\begin{equation}
\label{eq:maternextension}
	\mathbb{V} (\eta(x, t), \eta(y, s) ) = \lambda^2\ \Gamma(x,y) K(t, s),
\end{equation}
where $\Gamma$ refers to the pseudo empirical covariance structure obtained as explained in Section~\ref{sec:empiricalcovariance} and $K$ is the Mat\'ern-3/2 covariance function. 
The factor $\lambda$ allows us to incorporate an overall fluctuation amplitude of the 
ionosphere while maintaining the 
empirical correlation structure.
The full system can be cast into the following mixed model structure with hyper parameters
\begin{equation}
	d = F\beta + Z u + \epsilon,\quad \beta\sim\mbox{free mode},\quad u \sim {\cal N}( 0, \Gamma),\quad
    \epsilon \sim {\cal N}(0, \Sigma)
\end{equation}
where $d$ contains all observations at the observatories for some time interval, the coefficients $\beta$ are a vector of free modes (i.e.~with a diffuse prior) and $u$ encodes a correlated zero mean component, whereas
$\epsilon$ is the observational noise.

The observations are the available observatory data for the epochs we consider. 
Since not all observatories are
functional at all times we have to consider $d$ as an unstructured collection of (vector) 
observations at various
positions an times. It has dimension
$$
Dim (d) \leq 3 N_{Epoch} \times N_{Observatory}
$$
The free modes consist of the empirical mean, the four first EOFs, and for each observatory 
the constant offset. The first $5$ are individual for each epoch, whereas the latter is 
supposed to be a constant $3$ vector for each observatory for the epochs we consider.  
Thus $\beta$ has dimension 
$$
Dim(\beta) = 5 N_{Epoch} +  3 N_{observatory}.
$$
The random effects $u$ are the fluctuations around the free modes as described through the empirical covariance structure and the a priory covariance in time (see Eq.~\ref{eq:maternextension}). 
The dimension of the random vector $u$ depends on the truncation of our expansion into spherical harmonics
$$
Dim(u) = N_{Gauss} \times N_{Epoch}.
$$
The covariance of $u$ depends on an additional 
hyper-parameter, the global scaling $\lambda$ to take into account the overall variability 
 of the ionospheric field in the time range considered and we have
\begin{equation}
	\Gamma(x,t, y, s) = \lambda \Gamma_\text{iono}(x, t, y, s)
\end{equation}
The matrix $Z$ maps the Gauss coefficients to the corresponding field observations.
Therefore, the noise-free fluctuations of the ionospheric field at the observation points in time and space
around the "global" fixed effects have a covariance of
$$
\mathbb{V}(d |\beta, \epsilon) = Z \Gamma Z^T.
$$


The consistent treatment of $\beta$ as a component with a diffuse prior amounts to project the observations $d$ onto the space spanned by the rows of $F$. However, this projection has to be according to the geometry induced by the remaining ionospheric field component. Or to state it differently, the MAP estimator for $\beta$ minimizes the Mahalanobis distance induced by $\Gamma$ and $\Sigma$ to the data $d$. For fixed $\lambda$ we have
\begin{equation}
	\hat \beta = \mbox{argmin} (d-F\beta)^T (Z\Gamma Z^T + \Sigma)^{-1} (d-F\beta)~.
\end{equation}
Using a Cholesky factorization $Z\Gamma Z^T + \Sigma = L L^T$ we have equivalently
\begin{equation}
	\hat \beta = \mbox{argmin} \Vert L^{-1} d - L^{-1}F\beta \Vert~,
\end{equation}
which reduces the problem to a standard least square setting. Also the posterior distribution of $u$, for fixed
$\lambda$ can be obtained through linear algebra. 
For this and additional other details on the inversion technique can be found in \citet{Rasmussen:2006}. 
From the estimation of $\beta$ and $u$ we obtain an estimate of the ionosphere at the observatories and on all points of the Earth surface. 
\subsection{Modeling results}
\begin{figure}
	\begin{center}
		\includegraphics[width=\linewidth]{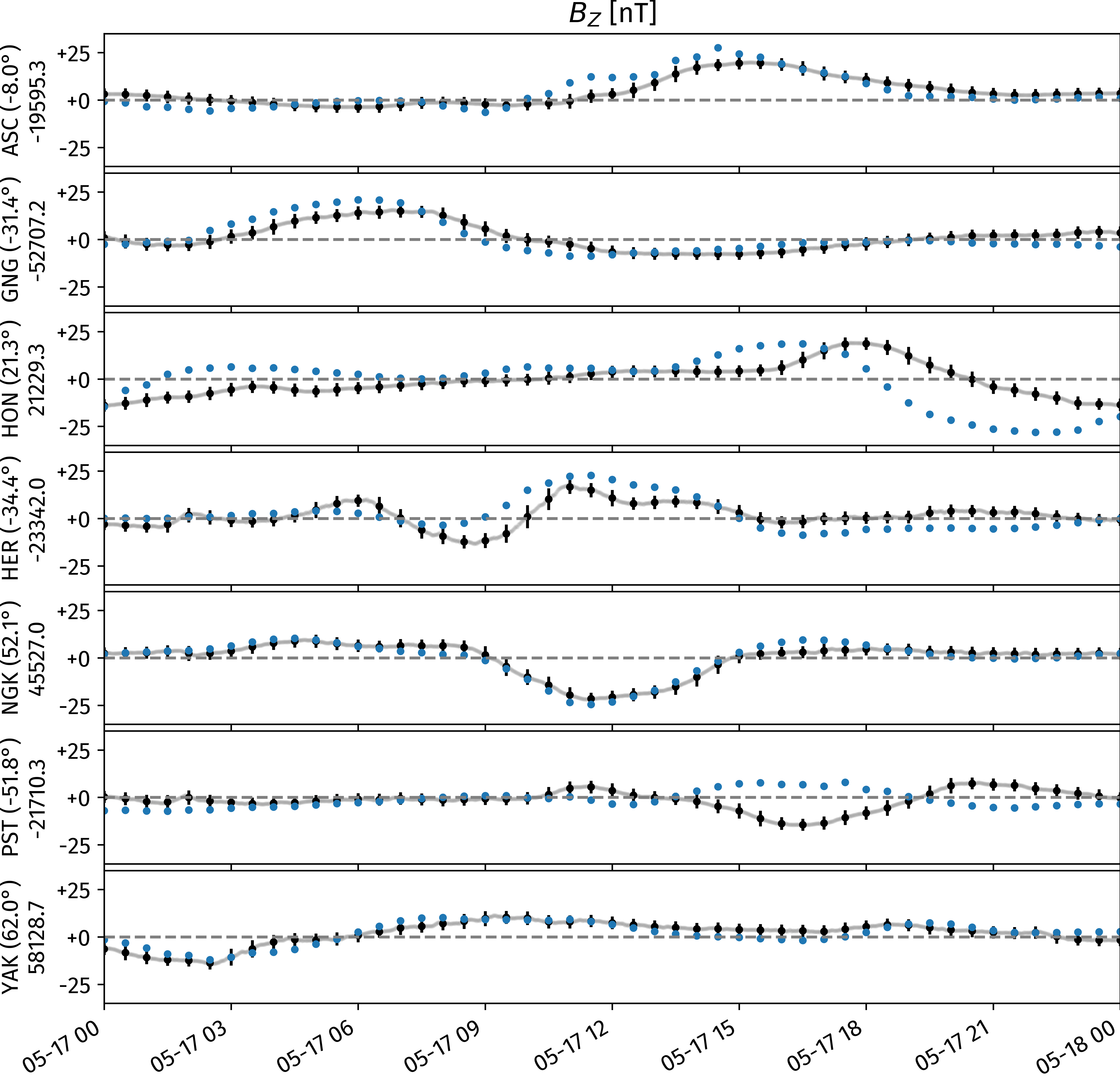}
		\caption{Time series of the downward component observed at different observatories. Data for the 17th of May 2014 was obtained from {INTERMAGNET} is shown as grey dots, together with 30 minute averages as black dots. The errorbars correspond to the 30 minute bin standard deviation with an additional error of 3 nT. The constant term reconstruction is shown as grey line and the full reconstruction as blue dots. The figures are vertically centered at the constant term. The observatory labels are given, together with the geographic latitude of the observatory and the value of the constant term reconstruction.}
		\label{fig:data}
	\end{center}
\end{figure}

For the 17th of May 2014, we obtained ground magnetic field observations from the {INTERMAGNET} network. The data type / quality was set to ``definitive''. To reduce the amount of data and to get rid of variations that are faster than the modeled dynamics, the reported data were averaged in 30 minute bins (see Figure~\ref{fig:data}). The standard deviation of the bins with an additional error of 3 nT was used to construct the noise covariance $\Sigma$.

\begin{figure}
	\begin{centering}
		\includegraphics[width=\linewidth]{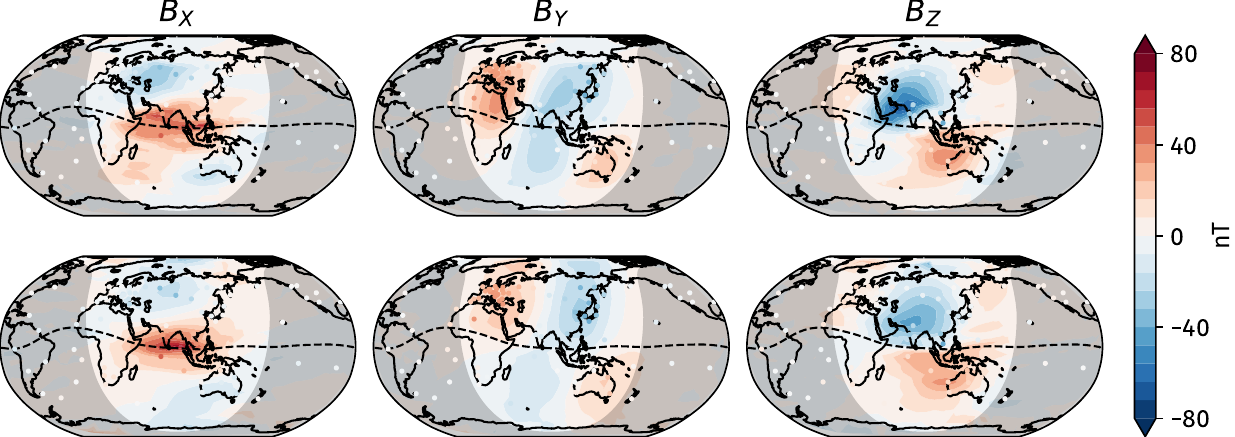}
	\end{centering}
	\caption{\textbf{Top row}: North ($X$), East ($Y$) and Down ($Z$) component of the magnetic field for the epoch 6 UTC, 17th of May 2014, as reconstructed from observatory data. \textbf{Bottom row}: DIFI prediction for the same epoch. The dots represent the observatories used in the inversion, the dot color refers to the observed magnetic field component at the observatory. The dashed line depicts the dip equator, derived from the Kalmag model \citep{Baerenzung:2020}.}
	\label{fig:outputs}
\end{figure}

\begin{figure}
	\begin{centering}
		\includegraphics[width=\linewidth]{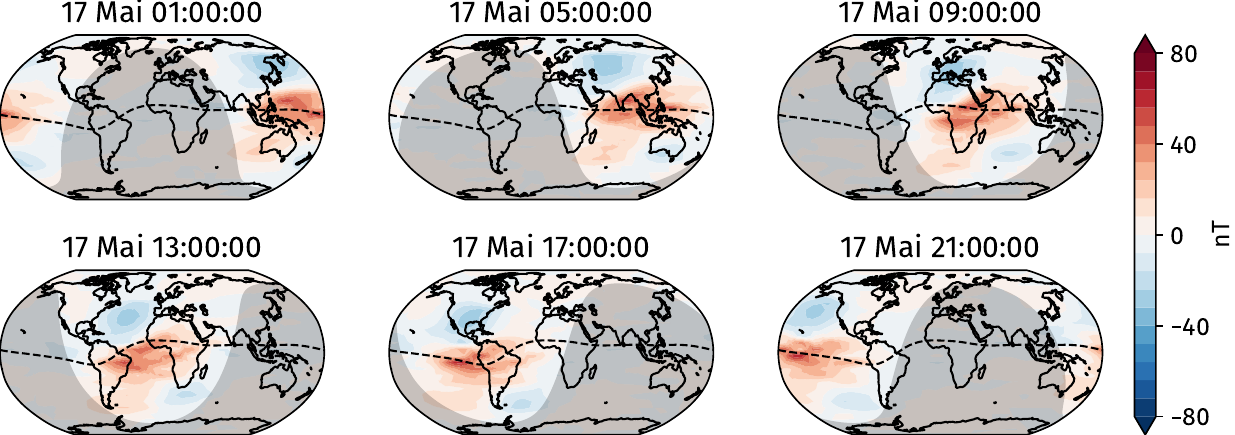}
	\end{centering}
	\caption{East component of the magnetic field for several epochs during the 17th of May 2014, as reconstructed from observatory data. The signature of the equatorial electrojet moving along the dip equator is clearly visible. The dashed line depicts the dip equator, derived from the Kalmag model \citep{Baerenzung:2020}. An animated version of this figure is available as supplementary material.}
	\label{fig:EEJ}
\end{figure}

From the set of global observatory data, the ionosphere was reconstructed on 30 minute knots over the whole day. Reconstruction of the local variability at various observatories is depicted in Figure~\ref{fig:data}. Global reconstructions of the ionosphere for several epochs during the day are depicted in  Figure~\ref{fig:outputs}, together with the DIFI prediction for the same epochs. Figure~\ref{fig:EEJ} shows the reconstructed signature of the equatorial electrojet for several epochs, an animated version of this figure is available with the supplementary material.

\section{Conclusion}\label{sec:conclusion}

We have developed a method to estimate the ionospheric field from observatory data. For this we have build a covariance structure using output from the empirical DIFI-7 model. This, together with Gaussian process based Bayesian inversion enables us to estimate the state of the ionosphere from ground based observations alone. The presented technique will possibly play a useful role, when assimilating satellite data and observatory data jointly. The currently used DIFI-7 model does not describe the polar field-aligned currents and is valid only between quasi-dipole latitudes -55$^\circ$ and 55$^\circ$. To include the field-aligned currents, a strategy similar to the one presented here could be applied utilizing the AMPS model \citep{Laundal:2018}, which is valid on the polar caps (above and below quasi dipole latitudes $\pm$60$^\circ$). A more independent covariance structure could be derived from numerical simulations of the ionosphere, if a Biot-Savart integrator is added to e.g.~the space wheather modeling framework (SWMF). This, together with more sophisticated treatment of the other field contributions may lead to further improved reconstruction of the ionospheric field and improved reconstruction and separation of the core, lithospheric and magnetospheric field. For the latter, a similar approach as demonstrated here is possible, using again output from numerical magnetohydrodynamics simulations. Our aim is to integrate this approach into the Kalman filter based inversion, presented in \citet{Baerenzung:2020}, in a future study.

\acknowledgments
V.L.~and M.H.~developed the concept as described in the project proposal submitted to the DFG (SPP1488).

\noindent K.F.~performed the analysis based on the UAM including all figures of the first versions taking into account the spatial covariance structure. 

\noindent After the UAM model became unavailable, M.S.~reproduced the study with the DIFI model and contributed the Mat\'ern time dependence.

\noindent K.F.~prepared the manuscript text with assistance from all co-authors. M.S.~provided all actual figures and adaptations to the DIFI model.

\noindent The DIFI model is publicly available (\url{https://geomag.colorado.edu/difi-7}). Observatory data used for the showcase inversion was obtained from INTERMAGNET (https://intermagnet.org/).

\noindent K.F.~acknowledges funding by the Deutsche Forschungsgemeinschaft (DFG, German Research Foundation), SPP1114

\noindent M.S.~acknowledges funding by the Deutsche Forschungsgemeinschaft (DFG, German Research Foundation), grant 388291411.

\bibliography{all}

\clearpage

\end{document}